\documentclass[12pt,onecolumn,peerreview,letterpaper]{IEEEtran}
\usepackage{amssymb}
\usepackage{amsmath}
\usepackage[dvips]{graphicx}
\usepackage{cite}
\usepackage{float}
\usepackage[T1]{fontenc}
\usepackage{array}
\usepackage{bm}
\usepackage[pdftex,bookmarks]{hyperref}

\newtheorem{lemma}{Lemma}
\newtheorem{remark}{Remark}

\newtheorem{example}{Example}

 \floatstyle{ruled}
\newfloat{algorithm}{tbp}{loa}
\floatname{algorithm}{Algorithm}
\begin{document}
\title{Design of Finite-Length Irregular Protograph Codes with Low Error Floors over the
Binary-Input AWGN Channel Using Cyclic Liftings}
\author{Reza ~Asvadi, Amir ~H. Banihashemi,~\IEEEmembership{Senior~Member,~IEEE,} and
Mahmoud~Ahmadian-Attari}
\author{Reza ~Asvadi\authorrefmark{1},
        Amir ~H. Banihashemi\authorrefmark{2},~\IEEEmembership{Senior~Member,~IEEE,}
        \\ and Mahmoud Ahmadian-Attari\authorrefmark{1}

\authorblockA{\authorrefmark{1}Department of Electrical and Computer Engineering\\
K. N. ~Toosi University of Technology, Tehran, Iran\\
 Emails: asvadi@ee.kntu.ac.ir, mahmoud@eetd.kntu.ac.ir}\\
\authorblockN{\authorrefmark{2}Department of Systems and Computer Engineering, Carleton
University,\\ Ottawa, Ontario, Canada\\
 Email: ahashemi@sce.carleton.ca}}

\maketitle
\begin{abstract}
We propose a technique to design finite-length irregular low-density
parity-check (LDPC) codes over the binary-input additive
white Gaussian noise (AWGN) channel with good performance in both the waterfall and
the error floor region. The design process starts from a protograph which
embodies a desirable degree distribution. This protograph is then lifted cyclically to
a certain block length of interest. The lift is designed carefully
to satisfy a certain approximate cycle extrinsic message degree (ACE)
spectrum. The target ACE spectrum is one with extremal properties,
implying a good error floor performance for the designed code.
The proposed construction results in quasi-cyclic codes which are attractive
in practice due to simple encoder and decoder implementation.
Simulation results are provided to demonstrate the effectiveness of the
proposed construction in comparison with similar existing constructions.
\end{abstract}


\section{INTRODUCTION}
Design of finite-length low-density parity-check
(LDPC) codes which perform well in both the waterfall and the error
floor regions is a challenging task. Irregular degree distributions
which are optimized to render superior waterfall performance will
often result in a high error floor in randomly constructed codes. To
improve the error floor performance of irregular codes, different
approaches have been examined. In one direction, iterative decoding
algorithms are manipulated to perform better in the error floor
region, see, e.g.,~\cite{HR-09}. In another direction, new
constructions of LDPC codes are
introduced~\cite{TJVW-04},~\cite{HEA-05},~\cite{Wang-06},~\cite{ICV-08},~\cite{VS-08},~\cite{VS-09},~\cite{JMSZ-09},~\cite{ABA-10}.
In this paper, we are interested in the latter approach. Our work is
closely related to the ideas and constructions introduced
in~\cite{TJVW-04},~\cite{VS-08}, and~\cite{VS-09}, on one hand, and
those in~\cite{Wang-06},~\cite{ICV-08}, and \cite{ABA-10}, on the
other hand.

The error floor performance of an LDPC code over the additive while Gaussian
noise (AWGN) channel is closely tied to graphical objects, referred to as
{\em trapping sets}~\cite{R-03}. A full characterization of dominant trapping sets
over the AWGN channel, particularly for irregular codes, is not available.
It is however known that enumerating such sets, in general, is a formidable
task~\cite{McGregor2007},~\cite{McGregor2008}.
Indirect measures of the error floor performance, which are computationally
more efficient, have thus been used in the design and the analysis of the
LDPC codes. It is well-known that cycles in the Tanner graph of the code
are responsible for the suboptimal performance of iterative decoding algorithms.
Different metrics have thus been introduced to measure the harmful effect of cycles.
The simplest such metric is the length of the shortest cycles in the graph,
called {\em girth}. Using this metric, Hu {\em et al.} proposed the {\em progressive edge-growth
(PEG) algorithm}~\cite{HEA-05}, which aims at maximizing the girth of the code's Tanner graph.
It was however observed that not all the cycles of the same length are
as harmful in iterative decoding~\cite{TJVW-04}. In~\cite{TJVW-04}, the
{\em approximate cycle extrinsic message degree (ACE)} of a cycle was introduced
as a metric to evaluate the harmfulness of a cycle. The larger the ACE, the less harmful a cycle
would be among a set of cycles of the same length. {\em ACE constrained} LDPC codes
were also designed in~\cite{TJVW-04} which outperformed random codes in the error floor region.
The ACE metric for a cycle was recently generalized to the {\em ACE spectrum} of
a Tanner graph in~\cite{VS-09}, where the authors also devised {\em generalized ACE constrained}
LDPC codes, further improving the error floor of the codes. Finally, in~\cite{VS-08},
PEG construction and generalized ACE constrained design were combined for even
superior error floor performance.

In~\cite{Wang-06}, {\em edge swapping} was proposed as a technique to increase the
{\em stopping distance} of an LDPC code, and thus to improve its error floor
performance over the binary erasure channel (BEC). Random
cyclic liftings was also studied in~\cite{Wang-06} and shown to improve
the average performance of the ensemble in the error floor
region compared to the base code. Ivkovic {\em et al.}~\cite{ICV-08} applied the
same technique of edge swapping between two copies of a base LDPC code to eliminate
the dominant trapping sets of the base code over the binary symmetric channel (BSC).
This was then generalized in~\cite{ABA-10} to cyclic liftings of higher degree, where
the liftings were designed to eliminate dominant trapping sets of the base code
by removing the short cycles which are part of the trapping sets.

In this work, starting from a protograph, we use cyclic liftings to construct LDPC codes.
The resulting codes are quasi-cyclic and thus attractive from implementation point of view.
Cyclic liftings are carefully designed to achieve a target ACE spectrum with extremal properties,
and thus the name {\em ACE constrained cyclic edge swapping} for the technique.
The extremal ACE spectrum would imply a good error floor performance for the code.

Compared to the constructions of~\cite{TJVW-04},~\cite{VS-08}, and~\cite{VS-09}, which do not
have any particular structure, the proposed construction is quasi-cyclic
and thus more implementation friendly. Moreover, the approach to achieve
a target ACE spectrum is different. In particular, in the proposed construction,
we first find a set of most vulnerable subgraphs of the protograph whose inverse image in the lifted
graph can potentially be small cycles with low ACE values. We then carefully assign
cyclic permutations to selected edges of these subgraphs such that their inverse
image satisfies the target ACE spectrum for the lifted graph.

In comparison with~\cite{ICV-08} and~\cite{ABA-10}, which start from rather large base codes (graphs),
we begin with a rather small base code.
This implies a more compact description for the code which can in turn
result in simpler encoder and decoder implementation. As a consequence of
a smaller base code, to achieve a given block length for the lifted code,
the lifting degrees in this work are larger compared to those in~\cite{ICV-08} and~\cite{ABA-10}.
Another difference is in the approach to design the cyclic liftings.
In~\cite{ICV-08} and~\cite{ABA-10}, the liftings are designed to eliminate dominant trapping sets of the base
code, implicitly assuming that such trapping sets are known and available. While this assumption
may be valid for hard-decision algorithms over the BSC, for the soft-decision algorithms
over the AWGN channel, the knowledge of dominant trapping sets, particularly for irregular codes,
is much harder to attain if not infeasible. Moreover,
the approach of~\cite{ICV-08} and~\cite{ABA-10}, can work only for the cases where
the base graph is large enough, and thus sparse enough, to allow for
the existence of a meaningful set of dominant trapping sets.\footnote{To explain this,
consider the $(155,64)$ Tanner code~\cite{TSF-01}, which has been used as a base code in constructions of
both~\cite{ICV-08} and~\cite{ABA-10}.
This code itself is a cyclic lifting of a complete bipartite graph $K_{3,5}$ with
5 variable nodes of degree 3 and 3 check nodes of degree 5. While one can
meaningfully find the dominant trapping sets of the $(155,64)$ Tanner code
for a certain decoding algorithm over a certain channel, the same is not possible for the
much smaller $K_{3,5}$ graph.}
This may not be the case for the small base graphs which are the subject of this work.

The remainder of this paper is organized as follows. Section
\ref{sec::prelimin} briefly reviews the definitions, notations
and concepts required in the rest of the paper.
The ACE constrained cyclic edge swapping method is presented in
Section \ref{sec::method}, followed by some simulation results in Section \ref{sec::result}.
The paper is then concluded in Section~\ref{sec::conclusion}.

\section{PRELIMINARIES: LDPC CODES, TANNER GRAPHS, CYCLIC LIFTINGS, AND ACE SPECTRUM}
\label{sec::prelimin}
\subsection{LDPC Codes and Tanner Graphs}
Consider a binary LDPC code ${\cal C}$ represented by a Tanner graph
$G = (V_b \cup V_c , E)$, where $V_b = \{b_1, \ldots, b_n\}$ and $V_c = \{c_1, \ldots, c_m\}$
are the sets of variable nodes and check nodes, respectively, and $E$ is the set of edges.
Corresponding to $G$, we have an $m \times n$ parity-check matrix
$H=[h_{ij}]$ of ${\cal C}$, where $h_{ij} = 1$ if and only if (iff) the
node $c_i \in V_c$ is connected to the node $b_j \in V_b$ in $G$;
or equivalently, iff $\{b_j,c_i\} \in E$. If all the nodes in the
set $V_b$ have the same degree $d_v$, and all the nodes
in the set $V_c$ have the same degree $d_c$, the corresponding
LDPC code is called  $(d_v,d_c)$-{\em regular}. Otherwise,
it is called {\em irregular}. For an irregular LDPC code, the
{\em degree distributions} of variable nodes and check nodes are
often described by the two polynomials, $\lambda(x)=\sum_{i=2}^{D_v}
\lambda_i x^{i-1}$ and $\rho(x)= \sum_{i=2}^{D_c} \rho_i x^{i-1}$,
respectively, where $D_v$ and $D_c$ are the maximum variable node degree
and the maximum check node degree, respectively, and $\lambda_i$ and $\rho_i$
are the percentage of the edges connected to the variable nodes and
check nodes of degree $i$, respectively. Alternatively, vectors
${\bm \lambda} = (\lambda_2,\ldots,\lambda_{D_v})$ and ${\bm \rho}=(\rho_2,\ldots,\rho_{D_c})$ can be
used for the description of the degree distributions. In this case,
the code is referred to as a $({\bm \lambda},{\bm \rho})$-{\em irregular} code.

A (directed) {\em walk} of length $n$ in a graph $G$ is a non-empty
alternating sequence $v_0e_1v_1\ldots v_{n-1}e_nv_n$ of nodes and
edges in $G$ such that $e_i=\{v_{i-1},v_i\}$ for all
$1\leq i\leq n$. If the two end nodes are the same, i.e., if $v_0=v_n$,
the walk is {\em closed}. A closed walk is {\em backtrackless}
if $e_i\ne e_{i+1}$ for $1\leq i\leq n-1$.
A backtrackless closed walk is called {\em tailless} if $e_n \ne e_1$.
We use the abbreviation TBC for tailless backtrackless closed walks.
A closed walk with distinct intermediate nodes is called a {\em cycle}.
The length of the shortest cycle(s) in the graph is called {\em girth}.
In bipartite graphs, including Tanner graphs, all closed walks have even
lengths, and thus the girth is an even number.

\subsection{Cyclic Liftings}
\label{subsec2.1}
Consider the cyclic subgroup $C_N$ of the {\em symmetric group} $S_N$ over the set of
integer numbers $Z_N \stackrel{\Delta}{=} \{0, \ldots , N-1\}$, with the group operation
defined as composition. The group $C_N$ consists of the $N$ circulant
permutations defined by $\pi_d(i) = i+d \:\;\mbox{mod}\:\; N, \: d \in Z_{N}$.
The permutation $\pi_d$ corresponds to $d$ cyclic shifts to the right.
Corresponding to $\pi_d$, we define the permutation matrix $I^{(d)}$
whose rows are obtained by cyclically shifting all the rows of the identity matrix $I_N$ by $d$
to the left. Clearly, $I^{(0)} = I_N$.
There is a natural isomorphism between (a) $C_N$, (b) the group of integers modulo $N$, $Z_N$, under addition,
and (c) the group of circulant permutation matrices under multiplication. This isomorphism
is defined by the correspondence between $\pi_d$, $d$ and $I^{(d)}$. In the following, we refer to
$d$ as the {\em shift} of the permutation $\pi_d$.

Consider the following construction of a graph $\tilde{G} = (\tilde{V},\tilde{E})$
from a graph $G = (V,E)$: We first make $N$ copies of $G$ such that for each node $v \in V$,
we have $N$ copies $\tilde{v} \stackrel{\Delta}{=} \{v^0 \ldots, v^{N-1}\}$ in $\tilde{V}$.
For each edge $e = \{u , v\} \in E$, we apply
a permutation $\pi^e \in C_N$ to the $N$ copies of $e$ in $\tilde{E}$ such that
an edge $\{u^i , v^j\}$ belongs to $\tilde{E}$ iff $\pi^e(i)=j$. The set
of these edges is denoted by $\tilde{e}$. The graph
$\tilde{G}$ is called a {\em cyclic} $N$-{\em cover} or a {\em cyclic} $N$-{\em lifting}
of $G$, and $G$ is referred to as the {\em base graph}, {\em protograph} or {\em projected graph}
corresponding to $\tilde{G}$.
We also call the application of a permutation $\pi^e \in C_N$ to the $N$ copies of $e$,
{\em cyclic edge swapping}, highlighting the fact that the cyclic permutation
swaps edges among the $N$ copies of the base graph.

In this work, $G$ is a Tanner graph, and we define the edge permutations from
the variable side to the check side, i.e., the set of edges $\tilde{e}$ in $\tilde{E}$ corresponding to
an edge $e = \{b,c\} \in E$ are defined by $\{b^i,c^{\pi^e(i)}\}, \: i \in Z_N$.
Equivalently, $\tilde{e}$ can be described by $\{b^{(\pi^e)^{-1}(j)},c^j\}, \: j \in Z_N$,
where $(\pi^e)^{-1}$ is the inverse of $\pi^e$ in $C_N$.
In other words, if $\pi_d \in C_N$ is a permutation from variable nodes to check nodes,
$\pi_{d'},\:d'= N-d\:\:\mbox{mod}\:\:N$, will be the corresponding permutation from check nodes
to variable nodes. (Note that $d'$ is the additive inverse of $d$ in $Z_N$.)
It is important to distinguish between the two cases when we
compose permutations on a directed walk.

To the lifted graph $\tilde{G}$, we associate an
LDPC code $\tilde{\cal{C}}$, referred to as the {\em lifted code},
such that the $mN \times nN$ parity-check matrix $\tilde{H}$ of $\tilde{\cal{C}}$
is equal to the biadjacency matrix of $\tilde{G}$. More specifically, $\tilde{H}$
consists of a total of $mn$ sub-matrices $[\tilde{H}]_{ij}\:, 1 \leq i \leq m, \: 1 \leq j \leq n$,
with each sub-matrix of dimension $N \times N$,
arranged in $m$ rows and $n$ columns. The
sub-matrix $[\tilde{H}]_{ij}$ in row $i$ and column $j$ is the circulant permutation matrix
corresponding to the edge $\{b_j,c_i\}$ when $h_{ij} \neq 0$;
otherwise, $[\tilde{H}]_{ij}$ is the all-zero matrix.
Let the $m \times n$ matrix $D=[d_{ij}]$ be defined
by $[\tilde{H}]_{ij} = I^{(d_{ij})},\:d_{ij} \in Z_N$ if $h_{ij} \neq 0$, and $d_{ij} = +\infty$, otherwise.
Matrix $D$, called the matrix of {\em edge permutation shifts}, fully describes $\tilde{H}$ and thus the
cyclically lifted code $\tilde{\cal{C}}$.

\subsection{ACE Spectrum and ACE Constrained Code Design}
Consider a cycle $\xi$ of length $\ell$ in a Tanner graph $G$.
The {\em approximate cycle extrinsic message degree (ACE)} of
$\xi$~\cite{TJVW-04} is defined as ACE($\xi\mbox{)} \stackrel{\Delta}{=} \sum_i (d_i-2)$, where $d_i$ is the degree
of the $i$th variable node of $\xi$, and the summation is over all the
$\ell/2$ variable nodes of $\xi$. Among cycles of the same length, those with larger
ACE values are less harmful to the performance of iterative decoding algorithms.
In~\cite{TJVW-04}, {\em ACE constrained LDPC codes} were designed whose Tanner graphs were free of short cycles
with small ACE values. These codes had much better error floors compared to similar random codes.
The same idea has been used in~\cite{DDJA-09} to devise cyclic lifts for protograph
LDPC codes.

In this paper, we extend the definition of ACE to TBC walks, i.e.,
for a TBC walk $W$ of length $\ell$ in a Tanner graph, the ACE is defined
as ACE($W\mbox{)} = \sum_i (d_i-2)$, where $d_i$ is the degree of the $i$th variable node of $W$, and
the summation is over all the $\ell/2$ variable nodes of $W$.

Given an LDPC code with a Tanner graph $G$, the ACE {\em spectrum of depth}
$d_{max}$ {\em of} $G$~\cite{VS-09} is defined as a $d_{max}$-tuple ${\bm \eta}(G) \stackrel{\Delta}{=}
(\eta_2,\ldots,\eta_{2d_{max}})$, where $\eta_{2i},\:i=1,\ldots,d_{max}$, is defined as
the minimum ACE value of all the cycles of length $2i$ in $G$.
Note that $\eta_{2i} = +\infty$, if there is no cycle of length $2i$ in $G$.
It is desirable to have larger ACE spectrum components for smaller cycles.
Given an ACE spectrum ${\bm \eta}=(\eta_2,\ldots,\eta_{2d_{max}})$ of depth $d_{max}$,
an LDPC code with a given Tanner graph $G$ (parity-check matrix $H$) is called $(d_{max},{\bm \eta})$
{\em ACE constrained} if for every value of $i, \:1 \leq i \leq d_{max}$, all
the cycles of length $2i$ in $G$ have an ACE value larger than or equal to
$\eta_{2i}$. In~\cite{VS-09}, a {\em generalized ACE constrained construction} of LDPC codes
was introduced, where given an ACE spectrum ${\bm \eta} = (\eta_2,\ldots,\eta_{2d_{max}})$
of depth $d_{max}$, a $(d_{max},{\bm \eta})$ ACE constrained LDPC code would be designed.
Such codes in general outperform the codes designed by the original ACE constrained
approach of~\cite{TJVW-04} in the error floor region. This is attributed to the fact that
the codes designed in \cite{TJVW-04} have a ``flat" ACE spectrum, while the ACE
spectrum for the codes constructed in~\cite{VS-09} do not have this limitation.
The generalized ACE constrained construction was also combined with the PEG construction in~\cite{VS-08}
to further improve the error floor.

Consider an ensemble $C^n({\bm \lambda},{\bm \rho})$ of LDPC codes with block length $n$ and degree distribution
$({\bm \lambda},{\bm \rho})$. A code ${\cal C}$ from this ensemble is said to have {\em extremal}
ACE spectrum properties~\cite{VS-09} if it has an ACE spectrum ${\bm \eta}=(\eta_2,\ldots,\eta_{2d_{max}})$,
with the property that there is no other code ${\cal C}'$ in $C^n({\bm \lambda},{\bm \rho})$
with an ACE spectrum ${\bm \eta}'=(\eta'_2,\ldots,\eta'_{2d_{max}}) \neq {\bm \eta}$, and
$\eta'_{2i} \geq \eta_{2i}$, for $1 \leq i \leq d_{max}$. Codes with extremal ACE spectrum properties
are expected to have good error floor performance~\cite{VS-09}. It is however a formidable task to
find LDPC codes with extremal ACE spectrum properties or to prove that a code has such properties.
Nevertheless, it would be desirable to aim at maximizing the ACE spectrum components in the code design process.
In this work, we perform this task systematically and in a greedy fashion,
focussing on one spectrum component at a time,
starting from the first ACE spectrum component $\eta_2$ followed by the rest of the components
in the increasing order of the cycle lengths until we reach $\eta_{2d_{max}}$.
We refer to the proposed scheme as
{\em ACE constrained cyclic lifting}, or {\em ACE constrained cyclic edge swapping}.

\section{ACE CONSTRAINED CYCLIC LIFTING}
\label{sec::method}
In our construction, we start from a given ACE spectrum
${\bm \eta}=(\eta_2,\ldots,\eta_{2d_{max}})$ of depth $d_{max}$,
and a given base code (graph) ${\cal C}$ of length $n$ with a certain degree distribution
$({\bm \lambda},{\bm \rho})$,
and design the non-infinity elements
of the matrix $D$ of edge permutation shifts such that the lifted code (graph)
is $(d_{max},{\bm \eta})$ ACE constrained. This results in
a $(d_{max},{\bm \eta})$ ACE constrained LDPC code $\tilde{\cal{C}}$
of length $nN$, where $N$ is the degree of the lifting. The lifted code
has the same degree distribution $({\bm \lambda},{\bm \rho})$
as the base code.

To design a lifting that satisfies an ACE spectrum ${\bm \eta}=(\eta_2,\ldots,\eta_{2d_{max}})$,
we need to eliminate all cycles of length $2i$ with ACE values less than $\eta_{2i}$
in the lifted graph. To eliminate these cycles in the lifted graph, all subgraphs of the
base graph whose inverse images include such low ACE cycles should be identified
and treated. The treatment is performed by the proper selection of the permutation shifts for the edges of
these problematic subgraphs of the base graph.

In the following, we show that the problematic subgraphs of the base graph
are short TBC walks with small ACE values.

\subsection{Images of Cycles in Cyclic Liftings}
\begin{lemma}
Consider a cyclic $N$-lifting $\tilde{G}$ of a Tanner graph $G$.
Consider a walk $W$ of length $\ell$ in $G$, which starts from a variable node
$b$ and ends at a variable node $b'$ with the sequence of edges
$e_1, \ldots, e_{\ell}$. Corresponding to the edges, we have the sequence
of permutation shifts $d_1,\ldots,d_\ell$. Then
the permutation shift that maps $\tilde{b}$, the inverse image of $b$ in $\tilde{G}$, to $\tilde{b'}$,
the inverse image of $b'$ in $\tilde{G}$, through the walk $\tilde{W}$ is $d$, where
\begin{equation}
d=\sum_{i=0}^{\ell-1} (-1)^i d_{i+1} \:\mbox{mod}\: N\:.
\label{eq2}
\end{equation}
\label{lem1}
\end{lemma}
\begin{proof} The proof is straightforward. See, e.g.,~\cite{Gross-Tucker-87}.
\end{proof}

The value of $d$ given in (\ref{eq2}) is called the {\em permutation shift} of the
walk from $b$ to $b'$. Clearly, the permutation shift of the walk from $b'$ to $b$ is equal to
$d' = N-d\:\:\mbox{mod}\:\:N$. If $b = b'$ and all the other nodes are distinct, then the walk
will become a cycle and depending on the direction of the cycle,
its permutation shift will be equal to $d$ or $d'$.

\begin{lemma}
Consider a cyclic $N$-lifting $\tilde{G}$ of a Tanner graph $G$.
Consider a TBC walk $W$ of length $\ell$ and ACE value $\eta$
in $G$. Suppose that the permutation
shift of $W$ is $d = 0$, and that there is no subgraph
of $W$ which is also a TBC walk with zero permutation shift.
Then, the inverse image of $W$ in $\tilde{G}$ consists of $N$
cycles, each of length $\ell$ and ACE value $\eta$.
\label{lemnb}
\end{lemma}
\begin{proof} Consider a variable node $b$ of the TBC walk $W$.
Consider the copy $b^i$ of $b$ in $\tilde{b}$ and follow
the walk in $\tilde{W}$ corresponding to $W$ starting from $b^i$. Since $d = 0$,
the walk ends at $b^i$ and is thus closed. Since $W$
has no subgraph which is also a TBC walk with zero permutation shift,
the above mentioned walk never goes through the same node twice.
With the same argument, it never meets $b^i$ again prior to the end of the walk.
The walk is thus a closed walk with distinct intermediate nodes and is therefore a cycle.
The same argument applies to all the $N$ copies of $b$ in $\tilde{b}$,
resulting in $N$ non-identical cycles in $\tilde{W}$. The ACE value
of each cycle is clearly the same as that of $W$, as the two have the same variable
degrees.
\end{proof}

\begin{lemma}
\label{thm::main} Any cycle of length $\ell$ and ACE $\eta$ in the cyclic
$N$-lifting $\tilde{G}$ of $G$ is projected onto a TBC
walk $W$ of length $\ell/k$ and ACE $\eta/k$ in $G$, where $k \geq 1$ is the
order of the permutation shifts of $W$.
\label{lemud}
\end{lemma}
\begin{proof} We first note that the two permutation shifts of $W$
corresponding to the two directions of $W$
are the inverse of each other in $Z_N$, and thus have the same order.
It is then simple to see that the image of a cycle $\xi$ must be a TBC walk $W$.
If the permutation shift $d$ of $W$ is zero, then $k = 1$, and
the result follows from Lemma~\ref{lemnb}. If $d \neq 0$, consider a node $b$
of $W$. Then with an argument similar to the one used in the proof of Lemma~\ref{lemnb},
one can see that the inverse image of $W$ starting from a node $b^i \in \tilde{b}$ of $\xi$
will return back to $b^i$ after traversing $k w$ edges of $\tilde{W}$, where $w$ is the
length of $W$. This implies that the length $\ell$ of the resulting cycle is $k w$.
The subgraph $\tilde{W}$ of $\tilde{G}$ in fact consists of $N/k$ cycles of
length $\ell$, all with the same degree distribution. The ACE result follows
readily from the above argument.
\end{proof}
For the cycles, as a special case of TBC walks, we have the following
result.
\begin{lemma}
Consider a cyclic $N$-lifting $\tilde{G}$ of a Tanner graph $G$.
Suppose that $\xi$ is a cycle of length $\ell$ and ACE $\eta$ in $G$. The inverse image of
$\xi$ in $\tilde{G}$ is then the union of $N/k$ cycles, each of length $k\ell$
and ACE $k\eta$, where $k$ is the order of the permutation shifts of $\xi$.
\label{lemvr}
\end{lemma}
\begin{proof} The proof for the number and the length of the cycles in the inverse image
of $\xi$ follows from the proof for the general case of permutation lifts given in
Theorem 2.4.3 of~\cite{Gross-Tucker-87}. The ACE results are simply a consequence of the
fact that each cycle $\xi'$ in the inverse image of $\xi$ has exactly the same variable degrees as $\xi$ does.
The multiplicity of the variable nodes of each degree in $\xi'$ however is $k$ times
that in $\xi$.
\end{proof}

In what follows, we refer to the value $k$ in Lemma~\ref{lemud},
as the {\em order} of the TBC walk, and use the notation
${\cal O}(W)$ to denote it. It is easy to see that
in a cyclic lifting of degree $N$, the order of a TBC walk $W$ is given by
\begin{equation}
{\cal O}(W) = \frac{N}{\mbox{gcd}(N,d)}\:,
\label{eqed}
\end{equation}
where $d$ is the permutation shift
corresponding to $W$, and gcd denotes the ``greatest common divisor.''

Based on Lemmas~\ref{lemnb}-~\ref{lemvr}, to eliminate short cycles with low ACE in the
lifted graph, we need to make sure that short TBC walks with low ACE in the base graph have
large orders.

\subsection{Structure of TBC Walks in the Base Graph}

Consider a base Tanner graph $G$.
Based on the results of the previous section, we are interested in
enumerating all the short TBC walks in $G$ and making sure that the edge permutation shifts
in these walks are selected such that the inverse images of the walks are
not short cycles with low ACE values. The following lemma is simple to
prove.
\begin{lemma}
A TBC walk in $G$ is either a cycle or consists of at least
two cycles.
\label{lemhd}
\end{lemma}

Based on Lemma~\ref{lemhd}, it is clear that the length of a TBC walk in $G$ is at
least $g$. All the TBC walks of length $g, \ldots, 2g -2$ are simple cycles.
TBC walks with length $\geq 2g$ consist of at least two cycles.

\begin{example}
\label{ex::ex1}
Consider the Tanner graph $G$ of Fig.~\ref{fig::tbcwalks} which has
3 variable nodes $b_1, b_2, b_3$ and 3 check nodes $c_1, c_2, c_3$.
The graph has two cycles of length 4 and one cycle of length 6 ($g=4$).
All three cycles are TBC walks. In addition, there are numerous other
TBC walks in $G$.
\begin{figure}[h]
\centering
\includegraphics[width = 0.4\textwidth]{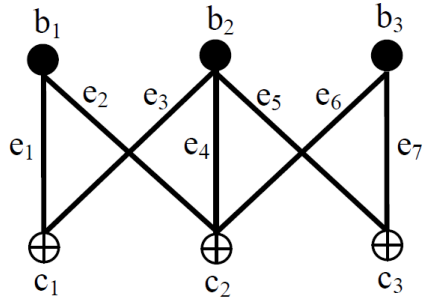}
\caption{A simple base Tanner graph} \label{fig::tbcwalks}
\end{figure}
One example is ${W}_1= b_1 e_2^+ c_2 e_4^- b_2 e_5^+ c_3 e_7^- b_3 e_6^+ c_2 e_4^- b_2 e_3^+ c_1 e_1^- b_1$,
which has length 8. Note that we have used superscripts $+$ and
$-$ to denote the direction of the edges in the TBC walk. In the rest of the paper,
we may only refer to a walk by its sequence of directed edges. For example,
$W_1$ can be represented as $e_2^+ e_4^- e_5^+ e_7^- e_6^+ e_4^-  e_3^+ e_1^-$.
Other examples of TBC walks in $G$ are ${W}_2=e_2^+ e_6^- e_7^+ e_5^- e_3^+ e_1^- e_2^+ e_4^- e_3^+ e_1^-$,
and ${W}_3= e_2^+ e_6^-  e_7^+ e_5^- e_4^+ e_6^- e_7^+ e_5^- e_3^+ e_1^-$, which both have length $10$.
Note that $W_1$ consists of two cycles of length 4, while each of $W_2$ and $W_3$
consists of a 4-cycle and a 6-cycle.
\end{example}

In order to find short TBC walks of the base graph, one can grow a
tree from every variable node in the graph, as the root, one layer
at a time and track the walks on the tree which pass through the
root node somewhere down in the tree. (Each layer of the tree is
constructed by first including all the check node neighbors of a
variable node except its parent node, and then by including all the
variable node neighbors of those check nodes except their parent
variable nodes.) By the construction, such a walk is a backtrackless
closed walk. One needs to only select those walks that are also
tailless. To find TBC walks of length at most $2l_{max}$, one needs
to grow the tree up to $l_{max}$ layers.\footnote{For Tanner graphs
with large variable/check degrees, to simplify the search for short
TBC walks, one can limit the search to cycles and subgraphs that
include two cycles.}

\subsection{ACE Constrained Cyclic Edge Swapping Algorithm}

Consider a target ACE spectrum
${\bm \eta}(\tilde{G})=(\eta_2,\eta_4,\ldots,\eta_{2d_{max}})$
for the cyclic $N$-lifting $\tilde{G}$ of the base graph $G$. Also,
consider a TBC walk $W$ of length $w$ and ACE $\eta(W)$ in $G$.
The TBC walk $W$ is referred to as (potentially) {\em problematic} if there exists a divisor $k$
of $N$ such that $k w \leq 2d_{max}$, and $k \eta(W) < \eta_{kw}$. Problematic
TBC walks are those whose inverse image in the lifted graph can be cycles that violate
the target ACE spectrum. One therefore should take proper care in assigning
permutation shifts to the edges of problematic TBC walks. In particular, if
the edge permutation shifts are assigned such that ${\cal O}(W) = k$,
where $k$ is the positive integer described above, then the $kw$-th component of
the target ACE spectrum will be violated by the inverse image of $W$.
The problematic TBC walks can be ordered based on
the comopnent of the target ACE spectrum that they would violate;
the smaller the index of the component, the more problematic the walk.
In the following, the ordered set of problematic TBC walks in G is denoted by ${\cal W}(G)$.
From this set, those that include edge $e$ are denoted by ${\cal W}^e(G)$.\footnote{A simpler
and still effective approach for ordering the TBC walks
in ${\cal W}(G)$ is to order them first based on their length, and then
for the walks of the same length, based on their ACE values.}

Forming an ordered set of problematic TBC walks ${\cal W}(G)$, the next step
is to go through this set, one TBC walk at a time and
assign proper permutation shifts to a selected set of edges from the chosen TBC walk such that
the inverse image of the walk does not violate the target ACE spectrum of the lifted graph.
In general, for a problematic TBC walk $W$, the policy is to select
the minimum number of edges of $W$ that can
make the inverse image of $W$ satisfy the ACE spectrum while maintaining
the satisfaction of the ACE spectrum by the previously processed problematic
TBC walks.

\begin{example}
\label{ex::ex2} Consider the base graph of Fig.~\ref{ex::ex1}.
Suppose that the goal is to satisfy the two ACE constraints $\eta_4 = \eta_6 = +\infty$
for the lifted graph.
The problematic TBC walks in this case are the 3 cycles
$\xi_1 = e_2^+ e_4^- e_3^+ e_1^-$, $\xi_2 = e_5^+ e_7^- e_6^+ e_4^-$,
and $\xi_3 = e_2^+ e_6^- e_7^+ e_5^- e_3^+ e_1^-$. To satisfy the ACE constraint
for the inverse images of these cycles, we need to make sure that
the order of all three cycles is larger
than one, i.e., the permutation shifts for all three cycles are nonzero.
Suppose that the
cycles are processed in the same order as listed above. To satisfy the ACE constraint
for the inverse image of $\xi_1$, we must have $d(\xi_1) \neq 0$. This can be satisfied by assigning
a nonzero permutation shift to only one edge of $\xi_1$.
For example, for $N=3$, one choice is $d(e_1) = 1$, which results in
$d(\xi_1) = d(e_2) - d(e_4) + d(e_3) - d(e_1)$ mod
$3 = 2 \neq 0$. Moving on to $\xi_2$, the inequality
$d(\xi_2) \neq 0$ can also be satisfied by assigning
a nonzero permutation shift to only one edge of $\xi_2$. To make sure that
this assignment will not affect $\xi_1$, we only search among the edges
of $\xi_2$ that do not belong to $\xi_1$. A proper choice can thus be, e.g.,
$d(e_5)=1$, which results in $d(\xi_2) = 1$. Finally, it is easy to see
that with the choices for the permutation shifts of $e_1$ and $e_5$,
we have $d(\xi_3)=1 \neq 0$, and thus no more edges need to be processed.
\end{example}

Given a set of problematic TBC walks ${\cal W}(G)$ of the base graph $G$, a
target ACE spectrum ${\bm \eta}(\tilde{G})=(\eta_2,\eta_4,\ldots,\eta_{2d_{max}})$
for the lifted graph $\tilde{G}$,
and the degree $N$ of the cyclic lifting, Algorithm~\ref{alg::des}
describes the proposed ACE constrained cyclic edge swapping.
At the output of Algorithm 1, we have the sets {\em SwappedSet} and {\em ShiftSet},
which contain the edges of the base graph that should be
swapped, and their corresponding permutation shifts, respectively.

\begin{algorithm}[h]
\textbf{Inputs:} A target ACE spectrum ${\bm \eta}(\tilde{G})=(+\infty,\eta_4,\ldots,\eta_{2d_{max}})$
of the lifted graph $\tilde{G}$, an ordered set of problematic TBC walks ${\cal W}(G)$ of
the base graph $G$, and the degree of cyclic lifting $N$.\\
1) Initialization: $ProcessedSet = \emptyset$, $SwappedSet = \emptyset$, $ShiftSet=\emptyset$.\\
2) Select the next problematic TBC walk $W \in {\cal W}(G)$. Denote
the length of $W$ by $w$ and its ACE value by $\eta(W)$.\\
3) $CurrentSet =$ edges of $W$.\\
4) $CandidateSet = CurrentSet  \setminus ProcessedSet$. \\
5) If $CandidateSet = \emptyset$, go to Step 8.\\
6) Select the edges ${\cal E}$ from the $CandidateSet$ that should be swapped, and assign their
permutation shifts ${\cal D}$ from $Z_N \setminus \{0\}$ such that $w {\cal O}(W) > 2d_{max}$ or
${\cal O}(W) \eta(W) \geq \eta_{w {\cal O}(W)}$. \\
7) $SwappedSet = SwappedSet \:\cup\: {\cal E}$, $ShiftSet = ShiftSet
\:\cup\: {\cal D}$,
and $ProcessedSet = ProcessedSet \:\cup\: CurrentSet$. Go to Step 12.\\
8) $CandidateSet = CurrentSet \setminus SwappedSet$. If $CandidateSet = \emptyset$, Stop.\\
9) Select an edge $e$ from $CandidateSet$
and assign a permutation shift $i \in Z_N \setminus \{0\}$ to it
such that the inverse images of all the TBC walks in ${\cal W}^e(ProcessedSet) \:\cup\:
W$ satisfy the ACE spectrum ${\bm \eta}(\tilde{G})$. If this is not feasible, go to Step 11.\\
10) $SwappedSet = SwappedSet \cup e$, $ShiftSet = ShiftSet \cup i$.
Go to Step 12.\\
11) $CandidateSet = CandidateSet \setminus e$, If $CandidateSet \neq
\emptyset$, go to Step 9. Else, stop.\\
12) If all the problematic TBC walks in ${\cal W}(G)$ are
processed, stop. Otherwise, go to Step 2. \caption{ACE Constrained Cyclic
Edge Swapping Algorithm} \label{alg::des}
\end{algorithm}

\begin{remark}
In Algorithm 1, the search for edges to be swapped and the permutation shift assignment to these edges
are performed in two phases. The first phase is in Steps 4 - 6, where any edge from previously processed
TBC walks is removed from the set of candidates for swapping. If the first phase fails, in that no edge
exists as a candidate for swapping ($CandidateSet = \emptyset$), then the algorithm switches to the second phase
in Steps 8 - 9, where only previously swapped edges are removed from the candidate set for swapping.
\label{remarkpd}
\end{remark}

\begin{remark}
To improve the performance of Algorithm 1, in Steps 6 and 9, it is advisable to
select edges that participate in a larger number of problematic TBC walks.
\label{remarkge}
\end{remark}

\begin{remark}
By increasing $N$, the size of the alphabet space for edge permutation shifts
increases. This in turn, allows for more problematic TBC walks to be accommodated.
In general, one can expect to achieve a better ACE spectrum as $N$ is increased.
\label{remarkke}
\end{remark}

In this work, we use Algorithm~1 in an iterative process to optimize the
ACE spectrum of the cyclic lifting. The goal is to achieve an ACE spectrum
of a certain depth $d_{max}$ with extremal properties. For this, we adopt a
greedy approach, where we attempt to maximize the ACE spectrum components,
one at a time, starting from $\eta_2$ and moving towards $\eta_{2d_{max}}$.
Ideally, we would like to achieve $\eta_{2i} = +\infty$, for $1 \leq i \leq d_{max}$,
but this is rarely possible for values of $d_{max}$ larger than 3.\footnote{One should
note that the greedy approach used in this work is not necessarily
the best approach in optimizing the ACE spectrum. In particular, there may be other
achieveable ACE spectrums for a given degree of lifting that result in a better
performance for the lifted code. Nevertheless,
our simulation results demonstrate that the greedy approach of this paper
is also quite effective in designing LDPC codes with good performance.}


\section{Numerical Results}
\label{sec::result}
In the following, two examples are presented where the proposed approach
is compared with the constructions of ~\cite{VS-08},~\cite{VS-09}, and~\cite{DDJA-09},
respectively.

\begin{example}
\label{exm::ex1} In this example, we consider the construction
of a binary irregular LDPC code with rate $\frac{1}{2}$,
and with the degree distribution similar to the one used in the examples of
\cite{VS-09}.\footnote{The degree distribution used in~\cite{VS-09}, which is optimized by density evolution,
is $\lambda(x)=0.2449x+0.20298x^2+0.00055x^3+0.1723x^4+0.37923x^{14}$
and $\rho(x)=x^7$.}. In~\cite{VS-09}, an LDPC code of length
$n=1000$ was designed using the generalized ACE constrained method.
This code had an ACE spectrum of $(+\infty,+\infty,16,9,5)$.

For a fair comparison, we would like to design an LDPC code with
a similar degree distribution and with a length of about $1000$ by using the
proposed method. Since our design is based on the cyclic lifting of a protograph,
we first need to design a rather small protograph with a degree distribution
close to that of~\cite{VS-09}.\footnote{It is important to note that there is a tradeoff in
the selection of the size of the protograph. On one hand, a small size is beneficial in having
a more compact description of the code and thus a simpler implementation
of the encoder and the decoder. On the other hand, it may be hard to implement
a given optimized degree distribution with high accuracy using a small protograph.
A smaller protograph also limits the number of variables that are available for
the optimization of the ACE spectrum of the lifted code.}
In this example, we design a protograph with parameters
$n=30$ and $m=15$ by the PEG algorithm. The degree distribution of
this protograph is
$\lambda(x)=0.2333x+0.2250x^2+0.1667x^4+0.3750x^{14}$, and
$\rho(x)=0.0583x^6+0.867x^7+0.0747x^8$. This base graph has ACE
spectrum $(+\infty,3,2,1,1)$.

We then apply the proposed method to design ACE constrained cyclic liftings of
various degrees for this protograph. The results are shown in Table \ref{table::table1} for
lifting degrees $N = 5, 10, 15, 20, 25$, and $30$.
It can be seen from Table~\ref{table::table1} that the ACE spectrum of the lifted graph
improves by increasing the lifting degree $N$.

\begin{table}[h]
\center
 \caption{ACE SPECTRUM OF CYCLIC LIFTINGS OF VARIOUS DEGREES FOR THE
$(30,15)$ BASE CODE OF EXAMPLE~\ref{exm::ex1}}
 \label{table::table1}
\begin{tabular}[c]{|c|c|}
\hline \centering {\em lifting degree (N) } & {\em ACE Spectrum }\\
\hline $5$ & $(+\infty,16,2,2,1)$ \\ \hline $10$ &
$(+\infty,26,2,2,1)$\\ \hline $15$ & $(+\infty,26,17,4,2)$ \\ \hline
$20$ & $(+\infty,+\infty,14,3,2)$ \\ \hline $25$ &
$(+\infty,+\infty,17,4,3)$\\ \hline $30$ & $(+\infty,+\infty,17,9,4)$\\
\hline
\end{tabular}
\end{table}

The closest block length to $1000$ is attained by the lifting degree $N=33$.
The lifted code in this case has length $n=990$, and we are able to achieve
the ACE spectrum $(+\infty,+\infty,17,10,5)$. This improves the ACE spectrum
of $(+\infty,+\infty,16,9,5)$ obtained in~\cite{VS-09} for $n = 1000$.

To have a more fair comparison with the constructions of~\cite{VS-08} and \cite{VS-09},
we also construct two codes with length $990$ whose degree
distributions are the same as our base graph by using the
generalized ACE constrained algorithm of \cite{VS-09} and
the generalized ACE constrained PEG algorithm of \cite{VS-08}.\footnote{More precisely,
the codes constructed by methods of~\cite{VS-08} and \cite{VS-09} have the same
variable degree distribution as the base graph, but their
check degree distributions are slightly different and depend
on the details of their design methods.} The constructed codes
by the methods of \cite{VS-09} and \cite{VS-08} have
ACE spectrums $(+\infty,+\infty,16,9,5)$ and $(+\infty,+\infty,17,9,4)$,
respectively, both inferior to the ACE spectrum of the code designed
by the ACE constrained cyclic lifting method. Moreover,
one should keep in mind that the proposed construction is quasi-cyclic and
thus more desirable for implementation.

The frame error rate (FER) and the bit error rate (BER) curves of
the three codes over the AWGN channel are presented in Fig. \ref{fig::FER1}.
The curves are for belief propagation decoding with
maximum number of iterations $100$. As can be seen, the code constructed
based on the proposed method outperforms the other codes, particularly
in the error floor region.

 \begin{figure}[h]
\centering
\includegraphics[width = 0.5\textwidth]{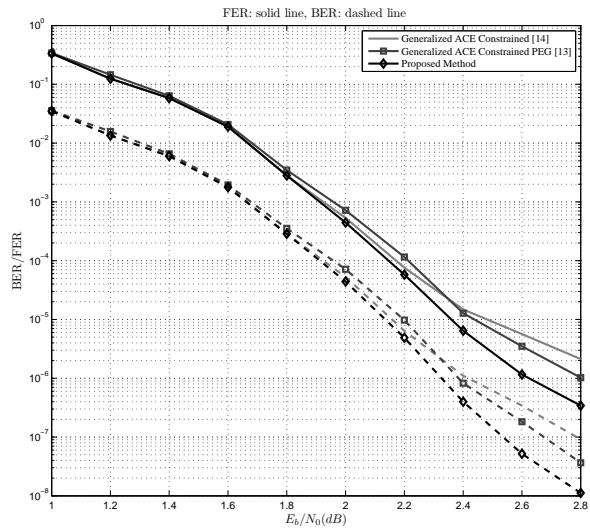}
\caption{BER/FER performance of the LDPC codes designed in
Example~\ref{exm::ex1}.} \label{fig::FER1}
\end{figure}

\end{example}

\begin{example}
\label{exm::ex2} As the second example, we consider the rate-compatible
protograph designed in~\cite{DDJA-09} for code rates
$\frac{1}{2}$, $\frac{5}{8}$, $\frac{3}{4}$, and $\frac{7}{8}$.
This protograph is shown in Fig.~\ref{fighd}. Different rates are
obtained by handling the non-trasmitted bits A, B, and C, differently.
If all the three bits are constrained to take a zero value, then
we obtain rate $\frac{1}{2}$. This is equivalent to removing the three
nodes and their edges from the protograph. Rate $\frac{5}{8}$ is obtained
if node A is an ordinary non-transmitted node with no bit assignment, but
nodes B and C are set to zero. Rate $\frac{3}{4}$
is obtained by setting only node C to zero. For rate $\frac{7}{8}$,
all three nodes are free of any bit assignment.

\begin{figure}[h]
\centering
\includegraphics[width = 0.4\textwidth]{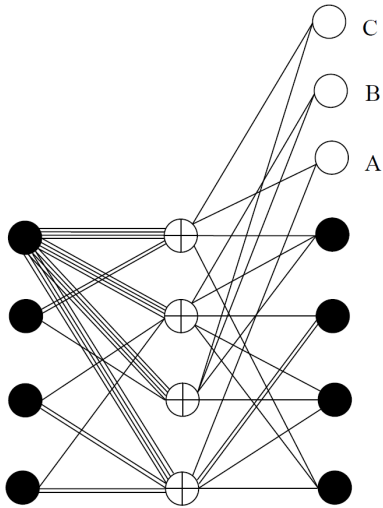}
\caption{Rate-Compatible protographs of rates $1/2$, $5/8$, $3/4$,
and $7/8$ used in Example~\ref{exm::ex2}.} \label{fighd}
\end{figure}

To remove the parallel edges, the protograph of Fig.~\ref{fighd} was lifted in \cite{DDJA-09}
by a lifting of degree 4 using the PEG algorithm. This graph, referred to hereafter as the base graph,
was then lifted by a cyclic lifting of degree 181 using the algorithm of~\cite{TJVW-04} to obtain
an LDPC code of length $n = 5792$. The ACE spectrums of the base graph and the lifted graph
of~\cite{DDJA-09} are $(+\infty,2,1,1,2)$ and $(+\infty,15,14,4,3)$,
respectively.

In Table \ref{table::table3}, the ACE spectrum of a number of cyclic liftings
of the base graph constructed by the proposed method is given. As expected,
the spectrum improves as the lifting degree increases.

\begin{table}[h]
\center
 \caption{ACE SPECTRUM OF CYCLIC LIFTINGS OF VARIOUS DEGREES DISCUSSED IN EXAMPLE~\ref{exm::ex2}}
 \label{table::table3}
\begin{tabular}[c]{|c|c|}
\hline \centering {\em lifting degree (N)} & {\em ACE Spectrum }\\
\hline $5$ & $(+\infty,15,2,3,2)$ \\ \hline $10$ &
$(+\infty,28,3,3,2)$\\ \hline $15$ & $(+\infty,28,16,2,2)$
\\ \hline
$20$ & $(+\infty,+\infty,15,3,2)$ \\ \hline $25$ &
$(+\infty,+\infty,16,3,3)$ \\ \hline $30$ & $(+\infty,+\infty,16,4,3)$\\
\hline
\end{tabular}
\end{table}

To compare with the code constructed in~\cite{DDJA-09}, we also design
a 181-lifting of the base code using the proposed method. The designed graph has
the ACE spectrum $(+\infty,+\infty,42,16,4)$, which significantly improves over
the ACE spectrum of the code designed in \cite{DDJA-09}.
We also compare the error rate performance of the designed code with
that of the code of~\cite{DDJA-09} over the AWGN channel in Fig.~\ref{fig::FER21}.
Belief propagation with a  maximum number of iterations $100$ is used
for decoding. Comparing the error rate performances of the two
codes, one can see that the designed code outperforms the code of \cite{DDJA-09}
across the whole range of rates, with significantly better performance
at the higher rates of $3/4$ and $7/8$, and particularly in the error floor region.

\begin{figure}[h]
\centering
\includegraphics[width = 0.5\textwidth]{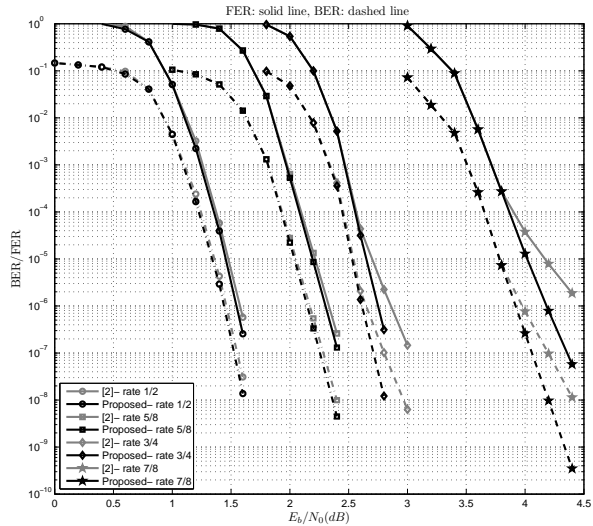}
\caption{BER/FER performance of rate compatible codes designed
in Example~\ref{exm::ex2}, and those of~\cite{DDJA-09}.} \label{fig::FER21}
\end{figure}

\end{example}

\section{CONCLUSION}
In this paper, we propose a method for the construction of finite-length irregular LDPC codes
with good waterfall and error floor performance. The constructed codes are quasi-cyclic protograph
codes and thus implementation-friendly. The performance of the codes are enhanced, particularly
in the error floor region, by the careful selection of edge permutation shifts for the vulnerable subgraphs of
the protograph. These subgraphs are the ones whose inverse image can be short cycles with low ACE values.
We demonstrate with a number of examples that the designed codes are superior to previously constructed codes
with similar parameters in both the ACE spectrum and the error correction performance.
\label{sec::conclusion}

\section*{Acknowledgment}
The authors wish to thank D. Divsalar, for providing them with the
codes constructed in~\cite{DDJA-09}, and D. Vukobratovi\'{c}, for
providing them with the codes constructed in ~\cite{VS-08}
and~\cite{VS-09}.

\end{document}